\documentclass[a4paper,11pt]{article}
\usepackage{pos}

\usepackage[utf8]{inputenc}
\usepackage{braket}

\newcommand{\rcite}[1]{\cite{#1}}

\newcommand{\refref}[1]{Ref.~\rcite{#1}}
\newcommand{\refrefs}[1]{Refs.~\rcite{#1}}
\newcommand{\eref}[1]{Eq.~(\ref{#1})}

\newcommand{\tref}[1]{Table~\ref{#1}}
\newcommand{\fref}[1]{Fig.~\ref{#1}}

\newcommand{\epow}[1]{\mathrm{e}^{#1}}

\title{Investigation of two-particle contributions to nucleon matrix elements}

\author[a,b]{Constantia Alexandrou}
\author[b]{Giannis Koutsou}
\author*[a]{Yan Li}
\author[c]{Marcus Petschlies}
\author[b]{Ferenc Pittler}

\affiliation[a]{Department of Physics,
University of Cyprus, P.O. Box 20537, 1678 Nicosia, Cyprus}

\affiliation[b]{Computation-based Science and Technology Research Center,
The Cyprus Institute, 20 Kavafi Str., Nicosia 2121, Cyprus}

\affiliation[c]{HISKP (Theory), Rheinische
Friedrich-Wilhelms-Universit{\"a}t Bonn, Nu{\ss}allee 14-16, 53115
Bonn, Germany}

\emailAdd{li.yan@ucy.ac.cy}

\abstract{
    We investigate contributions of excited states to nucleon matrix elements by studying the two- and three-point functions using nucleon and pion-nucleon interpolating fields. This study is carried out using twisted mass fermion ensembles with pion masses 346 MeV and 131 MeV. We compare the results obtained using these two ensembles and show preliminary results for nucleon charges.
}

\FullConference{The 40th International Symposium on Lattice Field Theory (Lattice 2023)\\
July 31st - August 4th, 2023\\
Fermi National Accelerator Laboratory\\}

\begin{document}
\maketitle

\section{Introduction}
Nucleon matrix elements $\braket{N|J|N}$ are important quantities to understand the structure of the nucleon. They can be extracted from the ratio of three- to two-point functions in the asymptotic limit:
\begin{align}
    \frac{\braket{0|O_N(t_{\text{sink}})J(t_{\text{ins}})\bar{O}_{N}(0)|0}}{\braket{0|O_N(t_{\text{sink}})\bar{O}_{N}(0)|0}} \xrightarrow{\text{$t_{\rm ins},\, t_{\rm sink}\, \rightarrow \infty$ }} \braket{N|J|N} \,.
\end{align}
The time separations used in practice, however, are limited and the ratio shows clear time dependence even at the largest available time separations. Since the time dependence comes from contributions of excited states, and the lowest excited state with the quantum numbers of the nucleon is a nucleon-pion state, we investigate the contribution of nucleon-pion states to the matrix elements.

\section{Generalized eigenvalue problem (GEVP) and its impact on three-point functions}
For a set of interpolating operators $\{O_i\}$ with fixed quantum numbers, one can compute the following matrix function:
\begin{align}
    C_{ij}(t) = \braket{0|O_i(t) O^\dagger_j(0)|0} \,.
\end{align}
The generalized eigenvalue problem (GEVP)
\begin{align}
    C_{ij}(t)\,v^n_j &= \lambda^n(t,t_0)\,C_{ij}(t_0)\,v^n_j 
\end{align}
can determine the eigenvalue $\lambda^n(t,t_0)$ and eigenvector $v^n_j$ of the $n$-th eigenmode. The eigenvectors should be independent of the reference time $t_0$.
Solving the GEVP one extracts excited state energies $E_n$ from eigenvalues $\lambda^n(t,t_0)$ via
\begin{align}
    \lambda^n(t,t_0) &= e^{-E_n (t-t_0)} \,+ \text{contributions from higher states}.
\end{align}
The GEVP also provides improved operators $O_n^\prime$ that increase the overlap with the $n$-th eigenstate by finding the optimal linear combination
\begin{align}
    O_n'\ket{0}\equiv v^n_j O^\dagger_j \ket{0} \propto \ket{n} \,.
\end{align}
Therefore, one can use the GEVP to construct an improved nucleon operator $O_N'$ to obtain a ratio of three- to two-point functions with reduced time dependence:
\begin{align}
    \frac{\braket{0|O_NJ\bar{O}_{N}|0}}{\braket{0|O_N\bar{O}_{N}|0}} \xrightarrow{\text{GEVP improved}} \frac{\braket{0|O_N'J\bar{O}_{N}'|0}}{\braket{0|O_N'\bar{O}_{N}'|0}} \,.
\end{align}

Here we consider a two-state model with $\ket{N}$ and $\ket{N\pi}$ to demonstrate how the GEVP works for three-point functions.
We start with two interpolating operators $O_N$ and $O_{N\pi}$. When acting on the vacuum they create states
\begin{align}\label{eq:unON}
    \bar O_N\ket{0}      &= \ket{N}+x\ket{N\pi} + \cdots \,,
    \nonumber\\
    \bar O_{N\pi}\ket{0} &= \ket{N\pi}+y\ket{N} + \cdots \,,
\end{align}
where without loss of generality we take one of the coefficients to by unity, and we also suppress the time-evolution factors to simplify the notations and the ellipsis denote higher excited states.
Then the improved nucleon operator $O_{N}'$ satisfies
\begin{align}
    \bar O_{N}'\ket{0} \equiv (\bar O_N - x\,\bar O_{N\pi}) \ket{0} &= (1-xy)\ket{N} \,,
\end{align}
and the improved three-point function $I$ can be expanded as
\begin{align}
    I&\equiv\braket{0|O_N'J\bar{O}_{N}'|0}=\braket{0|(O_N-x\,O_{N\pi})J(\bar{O}_N-x\,\bar{O}_{N\pi})|0} \nonumber\\
    &= \braket{0|O_N J \bar{O}_N|0} -x\braket{0|O_N J \bar{O}_{N\pi}|0} -x\braket{0|O_{N\pi} J \bar{O}_N|0} + x^2\braket{0|O_{N\pi} J \bar{O}_{N\pi}|0} \,.
\end{align}
Since the unimproved operators can both create two states as in \eref{eq:unON}, one can expand their three-point functions as follows
\begin{align}
    \braket{0|O_N J \bar{O}_N|0} &= \braket{N|J|N} + x\braket{N|J|N\pi} + x\braket{N\pi|J|N} + x^2\braket{N\pi|J|N\pi}  \nonumber\\
    \braket{0|O_N J \bar{O}_{N\pi}|0} &= y\braket{N|J|N} + \braket{N|J|N\pi} + xy\braket{N\pi|J|N} + x\braket{N\pi|J|N\pi}  \nonumber\\
    \braket{0|O_{N\pi} J \bar{O}_N|0} &= y\braket{N|J|N} + xy\braket{N|J|N\pi} + \braket{N\pi|J|N} + x\braket{N\pi|J|N\pi}  \nonumber\\
    \braket{0|O_{N\pi} J \bar{O}_{N\pi}|0} &= y^2\braket{N|J|N} + y\braket{N|J|N\pi} + y\braket{N\pi|J|N} + \braket{N\pi|J|N\pi}  \,.
\end{align}
Therefore, one has $16$ terms contributing to the improved three-point function as summarized in \tref{tab:GEVP3pt}.
\begin{table}[tbp]
    \caption{Decomposition of the improved three-point function $I=\braket{0|O_N'J\bar{O}_{N}'|0}$. Each contribution should be multiplied with the corresponding matrix element in the first row and the time-dependent exponential terms.}\label{tab:GEVP3pt}
    \centering
    \renewcommand\arraystretch{1.5}
    \begin{tabular}{lcccc} \hline\hline
        & $\braket{N|J|N}$ & ${\braket{N|J|N\pi}}$ & $\braket{N\pi|J|N}$ & $\braket{N\pi|J|N\pi}$ \\ \hline
        $I_1=\braket{0|O_N J \bar{O}_N|0}$ & $1$ & $x$ & $x$ & $x^2$ \\
        $I_2=-x\braket{0|O_N J \bar{O}_{N\pi}|0}$ & $-xy$ & $-x$ & $-x^2y$ & $-x^2$ \\
        $I_3=-x\braket{0|O_{N\pi} J \bar{O}_N|0}$ & $-xy$ & {$-x^2y$} & $-x$ & {$-x^2$} \\
        $I_4=x^2\braket{0|O_{N\pi} J \bar{O}_{N\pi}|0}$ & $x^2y^2$ & $x^2y$ & $x^2y$ & $x^2$ \\ \hline
        $I=I_1+I_2+I_3+I_4$ & $(1-xy)^2$ & $0$ & $0$ & $0$ \\
        $I^\prime=I_1+I_2+I_3$  & $1-2xy$ & $-x^2y$ & $-x^2y$ & $-x^2$ \\ \hline\hline
\end{tabular}
\end{table}
In \tref{tab:GEVP3pt}, elements from different columns should be multiplied with 
the matrix elements in the first row and the corresponding time-evolution factors, so for instance
given our 2-state model
\begin{align}
  I_3 &= -x\,\braket{0\,|\,O_{N\pi}\,\, J \,\, \bar{O}_N\, | \,0}
  \nonumber \\
  &=
  (-xy)\,\braket{N|J|N} \,\epow{-E_N\,t_{\mathrm{sink}} }
  \quad + \quad 
    (-x^2y) \,\braket{N|J|N\pi}\,\epow{-E_N\,(t_{\mathrm{sink}}- t_{\mathrm{ins}}) - E_{N\pi}\,t_{\mathrm{ins}}}
  \nonumber \\
  &\quad + (-x)\,\braket{N\pi|J|N} \,\epow{-E_{N\pi}\,(t_{\mathrm{sink}}- t_{\mathrm{ins}}) - E_{N}\,t_{\mathrm{ins}}}
  \quad + \quad 
    (x^2)\,\braket{N\pi|J|N\pi} \,\epow{-E_{N\pi}\,t_{\mathrm{sink}} }
  \label{eq:tab-example}
\end{align}
Thus elements from different columns in \tref{tab:GEVP3pt} cannot be compared directly. 
Elements from different rows, however, can be summed up as already done in the table. The sum of the 4 coefficients below each matrix element
in the first row then gives the total contribution of this matrix element to the improved three-point function $I = \braket{0| O'_N\,J\,\bar O'_N|0}$
in the 2-state model. Of course, the improved three-point function $I$ receives zero contributions from matrix elements with $\ket{N\pi}$ states,
which is borne out by the last-but-one row of the table.

Since $I_4$ is more difficult to compute, one can also look at the case $I^\prime=I_1+I_2+I_3$.
In this case, compared with the unimproved three-point function $I_1$, the contribution from ${\braket{N|J|N\pi}}$ or $\braket{N\pi|J|N}$ changes from $x$ to $-x^2y$, and the one from  $\braket{N\pi|J|N\pi}$ changes from $x^2$ to $-x^2$. If the contamination from the former is large, one should observe a significant change from $I_1$ to $I^\prime$ as long as $-xy$ is not close to $1$. 
Considering that $-xy\approx1$ would indicate $O_{N}\ket{0}=\ket{N}+x\ket{N\pi}$ and $O_{N\pi}\ket{0}\approx y\left(\ket{N}-x\ket{N\pi}\right)$ share a similar composition, it is unlikely to happen.
If the contamination from the latter is large, one should also observe a significant change because the sign of the contamination would flip.
Therefore, one should see a significant change if any of these $N\pi$ contamination terms are large. Contrary, if no change is observed, one can conclude that the contamination does not come from the $N\pi$ state under consideration. 

\section{Lattice setup}
We use two ensembles, one with a physical pion mass, the other one with a heavier pion mass, simulated using twisted mass clover-improved fermions as indicated in \tref{tab:ens}.

\begin{table}[tbp]
    \caption{Parameters of the ensemble used in this work. Further details are given in \refrefs{ETM:2015ned,Alexandrou:2018egz,ExtendedTwistedMass:2021gbo}. The right-most column gives the number of gauge configurations employed $N_{\text{cfg}}$.}\label{tab:ens}
    \centering
    \renewcommand\arraystretch{1.5}
    \begin{tabular}{cccccccc}
      \hline\hline
      Ensembles & Flavors & $N_L^3\times N_T$ & $a$ (fm) & $m_\pi$ (MeV) & $L$ (fm) & $m_\pi L$ & $N_{\text{cfg}}$ \\ \hline
      cA211.53.24 & 2+1+1 & $24^3 \times 48$ & 0.0947 & 346 & 2.27 & 3.99  & $2467$ \\
      cA2.09.48 & 2 & $48^3 \times 96$ & 0.0938 & 131 & 4.50 & 2.98  & $629$ \\ \hline\hline
    \end{tabular}
\end{table}

\begin{figure}[tbp]
    \centering
    \includegraphics[width=7cm]{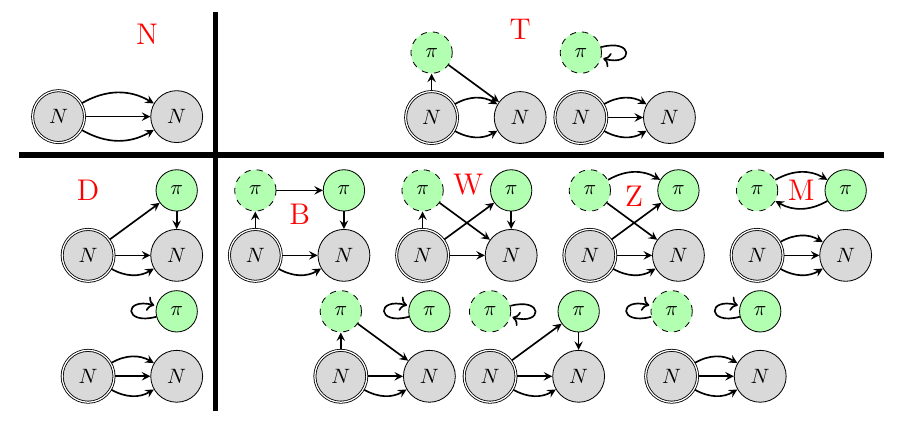}
    \includegraphics[width=7cm]{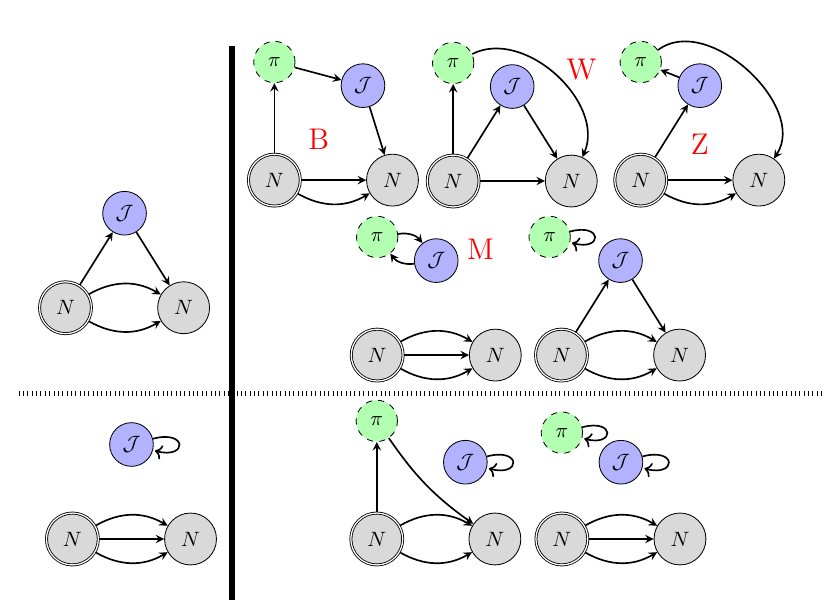}
    \caption{Different types of diagrams for two- (left) and three-point (right) functions under study.}\label{fig:23pt}
\end{figure}

We consider single nucleon and nucleon-pion operators at various momenta. 
We compute both the two- and three-point functions between them. For the three-point functions, $\braket{0|O_{N\pi}J\bar{O}_{N}|0}$ can be obtained via symmetry from $\braket{0|O_{N}J\bar{O}_{N\pi}|0}$, and $\braket{0|O_{N\pi}J\bar{O}_{N\pi}|0}$ has not been included in this work.
All the diagrams considered are summarized in \fref{fig:23pt}. Diagrams with pion loops are also included, 
because they are non-zero in twisted-mass lattice QCD, which breaks isospin symmetry. The three diagrams involving an insertion loop are still under investigation, and will not be included here.

Diagrams without loops were carried out in a similar way as in \refref{Alexandrou:2023elk}. For loops, we compute them stochastically with time dilution.

\section{Lattice results}

\begin{figure}[tbp]
    \centering
    \includegraphics[width=10cm]{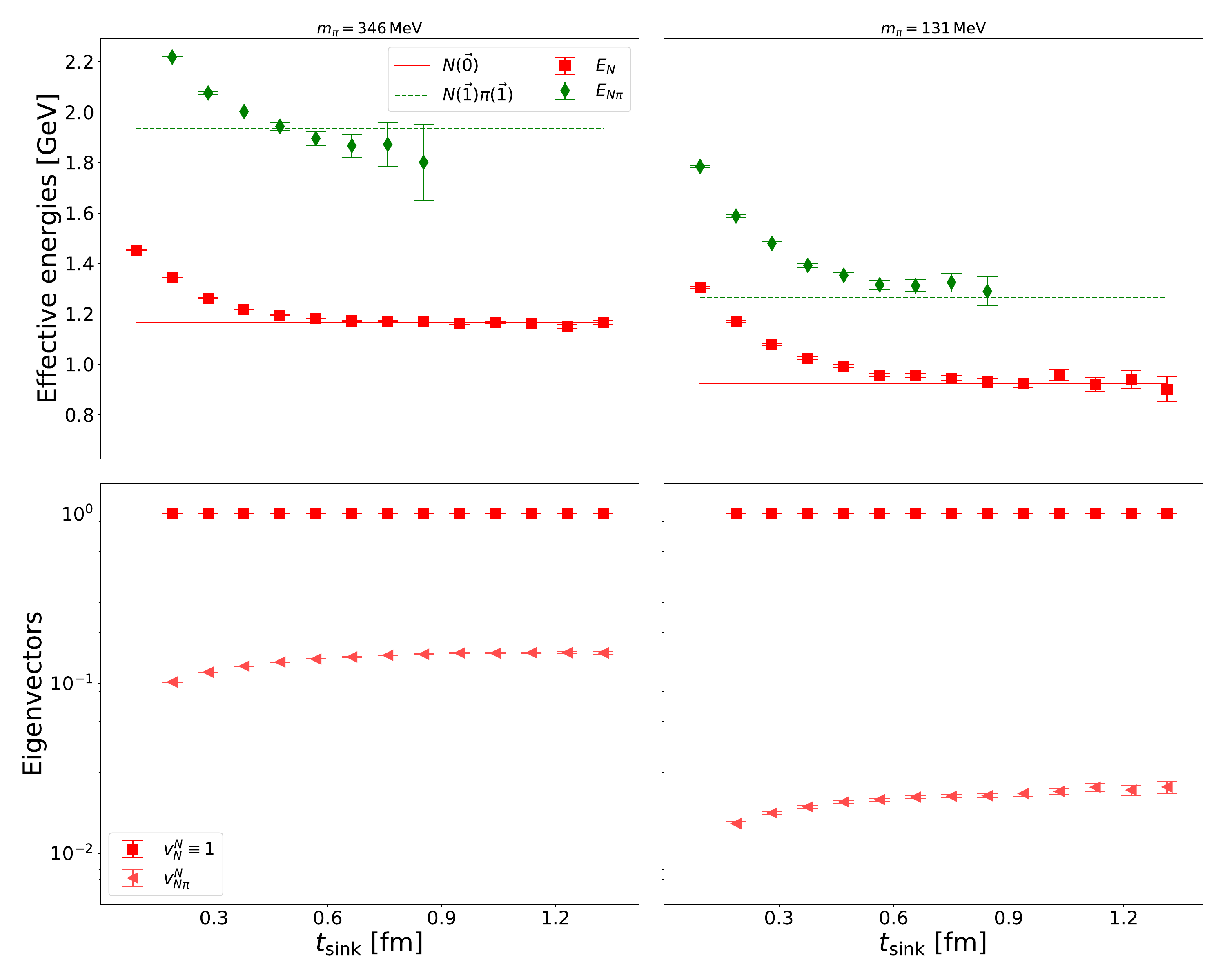}
    \caption{Effective energies and eigenvector components determined from the GEVP matrix of  two-point functions. Horizontal lines in the effective energy plots are for the nucleon and the non-interacting $N\pi$ levels. }\label{fig:2ptGEVP}
\end{figure}

In \fref{fig:2ptGEVP}, we show the results for the effective energies and eigenvector components determined by the GEVP. The effective energies show stability and asymptotically they yield values that are consistent with the nucleon and $\pi N$ non-interacting energies. The eigenvector components also show stability from around the same $t_{sink}$ as those for the effective energies.

\begin{figure}[tbp]
    \centering
    \includegraphics[width=9.2cm]{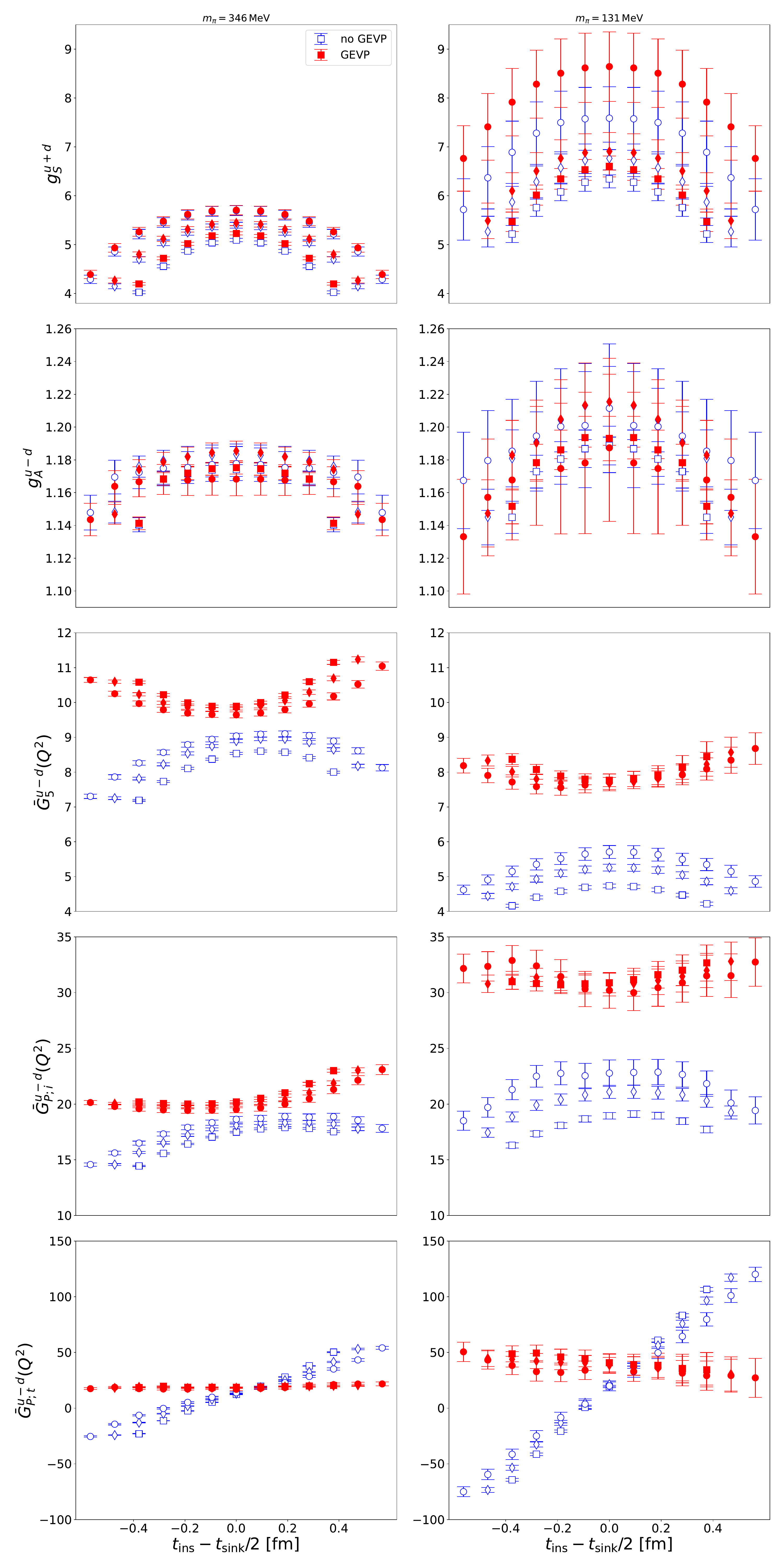}
    \caption{Ratios of three- to two-point functions before and after the GEVP improvement. Results are shown for $t_{\rm sink}=10a$ (square), $12a$ (diamond), $14a$ (circle) with different shapes. From top to bottom, the ratios are for the (connected) scalar  charge $g_S^{u+d}$, axial charge $g_A^{u-d}$, modified pseudoscalar form factor $\bar{G}_5^{u-d}(Q^2)$, and modified induced pseudoscalar form factors $\bar{G}_{P;i}^{u-d}(Q^2)$ and $\bar{G}_{P;t}^{u-d}(Q^2)$, where $Q^2$ corresponds to one lattice unit of momentum, $G_{P;i}$ and $G_{P;t}$ are extracted from the spacial and temporal axial matrix elements, respectively, and the modified form factors are related to the form factors via $\bar{G}_5=\frac{m_\pi^2+Q^2}{F_\pi m_\pi^2}m_qG_5$ and $\bar{G}_{P}=\frac{m_\pi^2+Q^2}{F_\pi m_\pi^2}\frac{Q^2}{4m_N}G_P$. We adopt the same conventions for these form factors as in \refref{Alexandrou:2023qbg}.}\label{fig:ratio}
\end{figure}

In \fref{fig:ratio}, we show the results for the ratio of three- to two-point functions before and after the GEVP improvement.
The upper panels are for zero transfer momentum $Q^2=0$. The first row is for the connected contribution to the $\pi N$-$\sigma$ term charge $g_S^{u+d}$. In this case, there are no significant changes between improved and unimproved matrix elements for both ensembles. Based on the discussion of the previous section,  we conclude that large excited state contamination in matrix element of the scalar operator from the lowest $\pi N$ state is unlikely.
For the axial charge $g_A^{u-d}$, shown in the panels in the  second upper row, we also do not observe significant changes. However, in this case, unlike for the matrix element of the scalar operator, no strong time-dependence is observed even before improvement using the GEVP.
We  also examine several combinations varying the current insertion and the momentum transfer.  While for most of them  we observe no significant changes, for a few cases there is improvement as discussed below.
We show results for the pseudoscalar $\bar{G}_5(Q^2)$ and induced pseudoscalar $\bar{G}_p$ form factors with pion pole removed in the third, fourth and fifth rows of \fref{fig:ratio} for one lattice unit of momentum transfer, $Q^2=0.283$ GeV$^2$ and $0.074$ GeV$^2$, for the CA211.53.24 and cA2.09.48 ensembles respectively. Each of these quantities shows a strong time dependence indicating high contribution from excited states. All of them show significant changes after applying the GEVP. Using the improved interpolating field, they show reduced time dependence, and in many cases, converge to a given constant  around  $t_{\rm ins}=t_{\rm sink}/2$. %
%
%
Most of the contamination is removed in particular for $\bar{G}_{p,t}(Q^2)$ computed using the axial vector in the temporal direction  showing indeed that $\pi N$ states are responsible for the contamination.

In those cases where a significant change is seen, there are five diagrams without an insertion loop for the $\braket{0|O_N J \bar{O}_{N\pi}|0}$ as shown in \fref{fig:23pt}. It turns out to be that the diagram labelled as `M' makes the dominant contribution. This was also observed in \refref{Barca:2022uhi}. It is also worthy noting that this diagram was observed to be enhanced only if the insertion operator can couple to the pion. This supports the chiral-perturbation-theory argument made by \refrefs{Bar:2018xyi,Bar:2019gfx}.

\section{Conclusions}
We use an improved nucleon interpolating field constructed from the GEVP to compute the three- to two-point function ratios for various nucleon form factors. 
We find that  excited state contamination in the scalar and axial charges matrix elements  is unlikely to come from the lowest $\pi N$ state, while for the pseudoscalar and induced pseudoscalar form factors a significant improvement is observed.

\section*{ACKNOWLEDGEMENTS}
We thank all members of the ETM collaboration for a most conducive cooperation.
We acknowledge computing time granted on Piz Daint at Centro Svizzero di Calcolo Scientifico (CSCS) via the project with id s1174, JUWELS Booster at the J\"{u}lich Supercomputing Centre (JSC) via the project with id pines, and Cyclone at the Cyprus institute (CYI) via the project with ids P061, P146 and pro22a10951.
Y.L. is supported by the Excellence Hub project "Unraveling the 3D parton structure of the nucleon with lattice QCD (3D-nucleon)" id EXCELLENCE/0421/0043 co-financed by the European Regional Development Fund and the Republic of Cyprus through the Research and Innovation Foundation.
F.P. acknowledges financial support by the Cyprus Research and
Innovation foundation under contracts with numbers
EXCELLENCE/0918/0129 and EXCELLENCE/0421/0195. 
C.A. and G. K. acknowledge partial support from the European Joint Doctorate AQTIVATE that received funding from the European Union’s research and innovation programme under the Marie Sklodowska-Curie Doctoral Networks action under the Grant Agreement No 101072344.
M.P. acknowledges support by the Sino-German collaborative research center CRC 110.

\bibliographystyle{JHEP}
\bibliography{refs}

\end{document}